\documentclass[prb,
superscriptaddress,showpacs,amsmath,amssymb]{revtex4}
\usepackage{amsfonts}
\usepackage{bm}
\usepackage{verbatim}

\begin{document}

\title{First law of de Sitter thermodynamics}

\author{G.E. Volovik}

\affiliation{Landau Institute for Theoretical Physics, 142432 Chernogolovka, Russia}

\date{\today}

\begin{abstract}
{The de Sitter state has a special symmetry: it is homogeneous, and its curvature is constant in space. Since all the points in the de Sitter space are equivalent, this state is described by
local thermodynamics. This state has the local temperature $T=H/\pi$ (which is twice the Gibbons-Hawking temperature), the local entropy density, the local energy density, and also the local gravitational degrees of freedom --  the scalar curvature ${\cal R}$ and the effective gravitational coupling $K$. On the other hand, there is the cosmological horizon, which can be also characterized by the thermodynamic relations. We consider the connections between the local thermodynamics and the thermodynamics of the cosmological horizon. In particular, there is the holographic connection between the entropy density integrated over the Hubble volume and the Gibbons-Hawking entropy of the horizon, $S_{\rm volume}=S_{\rm horizon}=A/4G$. We also consider the first law of thermodynamics in these two approaches. In the local thermodynamics, on the one hand, the first law is valid for an arbitrary volume $V$ of de Sitter space. On the other hand, the first law is also applicable to the thermodynamics of the horizon. In both cases, the temperature is the same. This consideration is extended to the contracting de Sitter with its negative entropy,  $S_{\rm volume}=S_{\rm horizon}=-A/4G$.
 }
\end{abstract}

\maketitle

 \tableofcontents



  \section{Introduction}
\label{IntroductionSec}

There are different approaches to thermodynamics and quantum theories of de Sitter space, see Refs. \cite{GibbonsHawking1977,Paddy2002,Odintsov2004,Markkanen2018,Akhmedov2020,Galante2023,Jacobson2023,Diakonov2025} and references therein.
The reason is that, unlike a black hole, whose thermodynamics is more or less generally accepted (see, however, Ref. \cite{Hooft2024}), the de Sitter state is not a localized object. It cannot be considered as a region limited by the cosmological horizon. The de Sitter state is the unbounded symmetric state with constant scalar curvature ${\cal R}$. All the points of the de Sitter space are equivalent, that suggests that the temperature of this state is uniform being the same for all the local (co-moving) observers. 

It is typically assumed that this temperature is associated with the temperature  of the Hawking radiation from the cosmological horizon -- the Gibbons-Hawking temperature $T_{\rm GH}=H/2\pi$, where $H$ is the Hubble parameter. However, if we consider the behaviour of any object, for example an atom placed in a de Sitter medium, it turns out that this object perceives this medium as a heat bath with double the Gibbons-Hawking temperature, $T=2T_{\rm GH}=H/\pi$. This coefficient of two provides the difference between two physical temperatures: the local temperature and temperature of Hawking radiation. This is one of the contradictions present in the construction of the thermodynamics of the de Sitter state.
Other contradictions concern the first law of thermodynamics; the sign of entropy, since in some approaches the entropy of the cosmological horizon is negative; and also the value and sign of the energy of the static patch.

Here we construct the consistent thermodynamics of the de Sitter state based on the local temperature $T=H/\pi$, which determines the entropy density, and on the contribution of the gravitational degrees of freedom to the thermodynamic pressure. In this approach, the presence of a cosmological horizon does not play any role. However, it turns out that the entropy of the region inside the cosmological horizon (the entropy $S_H$ of the Hubble volume) is positive and obeys the holographic bulk-surface correspondence: its value coincides with the Gibbons-Hawking entropy of the cosmological horizon, $S_H=A/4G$, where $A$ is the area of the horizon.
The first law of thermodynamics, related to the cosmological horizon, is also obtained.

This approach has been extended to the $f({\cal R})$ gravity and to thermodynamics of the contracting de Sitter, i.e. with $H<0$.
In this case the cosmological horizon has the property of the horizon of a white hole, and its entropy is negative, $S_H=-A/4G$. The local temperature of the contracting Universe is also negative, $T=-H/\pi$, so that the first law of thermodynamics, related to this  horizon, remains valid.

 \section{de Sitter local temperature and local entropy}
\label{HeatBathSec}

We consider the de Sitter thermodynamics using the Painlev\'e-Gullstrand (PG) coordinates,\cite{Painleve,Gullstrand} where the metric has no singularity at the cosmological horizon.
This allows us to consider local thermodynamics on the same basis at any point in de Sitter space, both inside and outside the horizon: all the points are equivalent. The PG metric is given by
 \begin{eqnarray}
ds^2= - c^2dt^2 +   (d{\bf r} - {\bf v}({\bf r})dt)^2 =
\nonumber
\\
= -c^2\left(1 -\frac{{\bf v}^2({\bf r})}{c^2}\right)-2{\bf v}({\bf r})\cdot d{\bf r}\,dt +d{\bf r}^2\,.
\label{G1}
\end{eqnarray}
Here ${\bf v}({\bf r})$ is the shift velocity (the velocity of the "vacuum"). In the de Sitter expansions, ${\bf v}({\bf r})=H{\bf r}$, where $H$ is the Hubble parameter, with $H>0$ corresponding to expansion and $H<0$ corresponding to contraction. The metric is (we use $c=1$):
 \begin{equation}
ds^2= - dt^2 +   (dr - Hr dt)^2+r^2 d\Omega^2 \,.
\label{G}
\end{equation}

Any external object in the expanding dS space radiates pairs of particles with the rate corresponding to the temperature $T$, which is twice the Gibbons-Hawking temperature, 
$T=2T_{\rm GH}=H/\pi$.\cite{Volovik2009}   
The behaviour of matter in the dS environment, such as spontaneous ionization of the hydrogen atom or proton decay,\cite{Volovik2025,Maxfield2022} also suggests that the de Sitter background has the temperature  $T=H/\pi$. For example the presence of a particle with mass $m$ generates the  process of triplication, $m \rightarrow 3m$, which is described by this temperature, $w\propto e^{-\Delta m/T}$, where $\Delta m=3m-m=2m$, see also Ref.\cite{Bros2008}. The created particles in turn triple and so on, creating an avalanche of particles in the de Sitter space.\cite{Polyakov2011} These processes lead to heating of matter and, accordingly, cooling of the thermostat and, thus, to relaxation of the vacuum energy,\cite{Volovik2024} solving the cosmological constant problems. For the problem of the vacuum energy relaxation,  it is important that the process of creation of matter by a matter immersed in a de Sitter environment has a higher exponent than the process of creation of matter from a vacuum by Hawking radiation, and also a significantly larger pre-exponential factor.\cite{Bros2008}  

Thus, instead of acting as an inversely populated medium as suggested by Polyakov,\cite{Polyakov2010} the de Sitter vacuum looks as a thermal bath with temperature $T=H/\pi$. With the local temperature  $T=H/\pi$, the energy density of the de Sitter state, which is the cosmological "constant" $\Lambda$, is:
\begin{equation}
 \epsilon_{\rm vac}\equiv \Lambda=\frac{3}{8\pi G}H^2=\frac{3\pi}{8 G}T^2\,.
\label{dSEnergyDensity}
\end{equation}
This determines the free energy density $F$ of the de Sitter state. From equation $F- T dF/dT=\epsilon_{\rm vac}$ one obtains $F(T)=-\epsilon_{\rm vac}(T)$, and thus the entropy density $s_{\rm dS}$ is:
\begin{equation}
s_{\rm dS}= - \frac{\partial F}{\partial T} =\frac{3\pi}{4G}T=\frac{3}{4G}H\,.
\label{dSEntropyDensity}
\end{equation}

Since the dS state is homogeneous, i.e. its local thermodynamic variables do not depend on position, we can consider thermodynamic relations for an arbitrary volume $V$, regardless of whether it is smaller or larger than the volume of space inside the horizon. The energy and entropy of the volume $V$ are:
  \begin{equation}
S_V=V s_{\rm dS} \,\,,\,\, E_V= V \epsilon_{\rm vac} \,.
\label{dSEntropyV}
\end{equation}

The entropy $S_H$ of the part of the de Sitter space, which is surrounded by the cosmological horizon, i.e. the total entropy of the Hubble volume $V_H=(4\pi/3) \,r_H^3$, is:
  \begin{equation}
S_{\rm Hubble}=V_H s_{\rm dS}= \frac{A}{4G}\,,
\label{dSEntropyHubble}
\end{equation}
where $A=4\pi r_H^2$ is the horizon area and $r_H=1/H$. 

The bulk entropy in Eq.(\ref{dSEntropyHubble}) exactly coincides with the entropy of the cosmological horizon suggested by Gibbons and Hawking. However, this global entropy comes from the local entropy of the de Sitter state, rather than from the horizon degrees of freedom. This demonstrates the holographic bulk-surface correspondence.

 \section{Thermodynamics modified by gravity}
\label{HeatBathSec}

The conventional Gibbs-Duhem law in the non-gravitating system is:
 \begin{equation}
T s=\epsilon + p \,.
\label{CoventionalFirstLaw}
\end{equation}

When gravity is included the thermodynamically conjugate variables are added, $K$ and ${\cal R}$,\cite{Volovik2024,Volovik2024c} which describe the gravitational degrees freedom. Then we have the modified thermodynamics law:
\begin{equation}
T s=\epsilon + p+ K{\cal R} \,.
\label{FirstLawGravity}
\end{equation}
In the Einstein theory $K=1/16\pi G$.
For the de Sitter state, the similar $K$ can be introduced in any theory expressed in terms of the general curvature tensors. This is because of the special symmetry of the de Sitter state, where all the curvature tensors can be expressed in terms of the scalar curvature ${\cal R}$.
 
It is convenient to introduce the modified pressure $P= p+ K{\cal R}$, then one returns to the conventional Gibbs-Duhem relation, but with pressure $P$: 
 \begin{equation}
T s=\epsilon +P \,,
\label{FirstLaw}
\end{equation}
and one obtains the first law of thermodynamics in the form
\begin{equation}
TdS = dE + PdV\,.
\label{TdS}
\end{equation}

 In the de Sitter state in Einstein gravity with the cosmological constant in Eq.(\ref{dSEnergyDensity}) one has:
 \begin{equation}
P =p_{\rm vac} + K{\cal R}=-p_{\rm vac} =\epsilon_{\rm vac} = \frac{3H^2}{8\pi G} = \frac{3\pi T^2}{8G}\,.
\label{Modified}
\end{equation}
 The equation of state in terms of the modified pressure is 
  \begin{equation}
P =w\epsilon_{\rm vac} \,\,,\,\, w=1\,.
\label{Zeldovich}
\end{equation}
  This is equivalent to the equation of state for the Zel'dovich stiff matter,\cite{Zeldovich1962} in which the speed of sound $c_s$ is equal to the speed of light, $c_s^2=dP /d\epsilon_{\rm vac}=1$.
 
  \section{Thermodynamics in  $f({\cal R})$ gravity}
\label{ExtensionSec}
 
  In the $f({\cal R})$ gravity the action is
 \begin{equation}
W=\int d^4x \sqrt{-g}f({\cal R}) + W_{\rm m} \,,
\label{fRaction}
\end{equation}
where $W_{\rm m} $ is the action for conventional matter.
Here, the cosmological constant (or vacuum energy) is included into $f({\cal R})$.
In the Einstein limit
\begin{equation}
f({\cal R}) =K {\cal R} - \epsilon_{\rm vac}  \,,
\label{Einstein}
\end{equation}
and $K=1/16\pi G$. 

In the $f({\cal R})$ theory the variable $K$ is the thermodynamically conjugate to the scalar curvature,  $K=df/d{\cal R}$. In the de Sitter state the equilibrium value of the curvature is given be equation
\begin{equation}
2f({\cal R})={\cal R}\frac{df}{d{\cal R}} \,,
\label{Equilibrium}
\end{equation}
and thus the equilibrium value of the gravitational coupling in the expansion with the Hubble parameter $H$ is
\begin{equation}
K=\frac{df}{d{\cal R}}\bigg|_{{\cal R}=12H^2}\,.
\label{GravitationalConstant}
\end{equation}
This $K$ determines the effective Newton constant $G=1/16\pi K$ which enters the
 entropy of the Hubble volume in Eq. (\ref{dSEntropyHubble}). 
 
 The equilibrium value of curvature also determines the vacuum energy (the cosmological "constant"). The role of the dark energy $\epsilon_{\rm vac}$ is played by the equilibrium value of $f(\cal R)$, i.e.
 $\epsilon_{\rm vac}(H)=f({\cal R}=12H^2)$. This means that in general the cosmological "constant" does not necessarily enter explicitly into $f(\cal R)$, but emerges in the $f(\cal R)$ gravity. In this description, the cosmological "constant" is not a constant, but represents the part of the gravitational degrees of freedom that relaxes due to interactions with matter.
 
All this shows that the first law in Eq. (\ref{TdS}) works for the de Sitter state also in $f(\cal R)$ gravity. Moreover, it applies to de Sitter in any gravity, if it is expressed in terms of curvature tensors. This is because, due to de Sitter symmetry, all curvature tensors can be expressed in terms of scalar curvature.
This is also the reason why the scalar curvature ${\cal R}$ and its conjugate $K$ serve as thermodynamic variables. This is similar to a pair of non-extensive electrodynamic variables in the thermodynamics of dielectrics: the electric field ${\bf E}$ and the electric induction ${\bf D}$, see also Ref.\cite{KlinkhamerVolovik2008}. Being non-extensive gravitational variables, $K$ and ${\cal R}$ naturally enter into the modified pressure, $P= p_{\rm vac}+ K{\cal R}$.
 
 \section{First law of thermodynamics for cosmological horizon}
\label{HorizonSec}

 The first law of thermodynamics in Eq.(\ref{TdS}) can also be applied to the part of the de Sitter space inside the cosmological horizon -- to the Hubble volume $V_H$:
  \begin{equation}
TdS_H = dE_H + P  dV_H \,.
\label{FirstLaw1}
\end{equation}
Indeed, for $T=H/\pi$ one has:
  \begin{eqnarray}
TdS_H =T\frac{dA}{4G}=\frac{H}{\pi}d \left(\frac{\pi}{H^2G} \right)=-2\frac{dH}{GH^2}\,,
\label{dSH}
\\
dE_H= d\left(\frac{4\pi}{3H^3}\frac{3H^2}{8\pi G} \right) = d\left(\frac{1}{2GH}\right)
=-\frac{1}{2}\frac{dH}{GH^2}\,,
\label{dEH}
\\
P  dV_H= \frac{3H^2}{8\pi G} d\left(\frac{4\pi}{3H^3}\right) = -\frac{3}{2}\frac{dH}{GH^2} \,,
\label{dVH}
\end{eqnarray}
and Eq.(\ref{FirstLaw1}) is satisfied:
\begin{equation}
-2\frac{dH}{GH^2} =-\frac{1}{2}\frac{dH}{GH^2} -\frac{3}{2}\frac{dH}{GH^2} \,.
\label{FirstLaw2}
\end{equation}

As follows from Eq.(\ref{dEH}), the energy of the Hubble volume in this approach is: 
\begin{equation}
E_H=\frac{1}{2GH}\,.
\label{EnergyHubble}
\end{equation} 

Note that the proper energy $E_H$ of the de Sitter Universe (actually the energy of the so-called static patch) is not well determined in the literature, and it is different for different approaches. There were different suggestions such as $E_H=1/(2GH)$, $E_H=-1/(2GH)$, $E_H=1/(GH)$, etc., see e.g. Section 2.1 in Ref. \cite{Paddy2002}. The zero value of the energy is also discussed, see e.g. Ref. \cite{Volovik2024d}. 
The same concerns the first law of thermodynamics.\cite{Diakonov2025,Odintsov2025} 
 For example, if one takes the negative value of energy, $E_H=-1/(2GH)$, and the Gibbons-Hawking temperature, $T=T_{\rm GH}=H/2\pi$, the first law in Eq.(\ref{FirstLaw1}) will be also satisfied (see also Ref. \cite{Odintsov2025a}):
\begin{equation}
-\frac{dH}{GH^2} =\frac{1}{2}\frac{dH}{GH^2} -\frac{3}{2}\frac{dH}{GH^2} \,.
\label{FirstLaw3}
\end{equation} 

Some approaches assume the negative entropy of the cosmological horizon,\cite{Jacobson2023,Diakonov2025} which we consider in the next section.

 \section{Negative entropy of contracting de Sitter}
\label{NegativeSec}

In Ref. \cite{Diakonov2025} another realization of the first law in Eq.(\ref{FirstLaw1}) was suggested. In this scenario the following values of thermodynamic variables were used:
\begin{equation}
E_H=\frac{1}{2GH} \,,\,T=\frac{H}{2\pi}\,,P=p_{\rm vac} \,,\,  S_H= - \frac{A}{4G}\,.
\label{setDiak}
\end{equation} 
With this choice, the first law is obeyed (see Eq.(3.22) in Ref. \cite{Diakonov2025}):
\begin{eqnarray}
T_{\rm GH}dS_H = dE_H + P dV_H \,,
\label{FirstLawDiakonov}
\\
\frac{dH}{GH^2} =-\frac{1}{2}\frac{dH}{GH^2} +\frac{3}{2}\frac{dH}{GH^2} \,.
\label{FirstLaw3}
\end{eqnarray} 
It is valid only under assumption that the entropy of the cosmological horizon is negative, $S_H=-A/4G$.

Both the Gibbons-Hawking temperature and the negative entropy are inconsistent with the approach in section \ref{HorizonSec}. However, negative entropy itself is not that unusual. Negative entropy has been discussed in Ref. \cite{Odintsov2002}.
Negative entropy occurs also for a white hole.\cite{Volovik2020} The rate of the macroscopic quantum tunneling from the black hole to the white hole of the same mass $M$ demonstrates that entropy of the white hole is with minus sign the entropy of the black hole:
\begin{equation}
S_{\rm wh}(M)=-S_{\rm bh}(M)=-\frac{A}{4G} \,.
\label{whEntropy}
\end{equation} 

The negative entropy can be obtained in our approach too, but only if we consider the contracting de Sitter state, instead of the expanding one. The contacting state has the negative Hubble parameter, $H<0$, in Eq.(\ref{G}) and thus it is natural to assume that the corresponding temperature is negative, $T=H/\pi <0$. If so, then from Eq.(\ref{FirstLaw1}) it follows that the entropy of the Hubble volume should be also negative, $S_H=-A/4G$. This is justified because the cosmological horizon in a contracting de Sitter has the same properties as the horizon of a white hole - a distant observer sees that matter cannot cross the horizon, although matter escapes from it. Therefore such cosmological horizon must also have negative entropy as in Eq.(\ref{whEntropy}).

The interplay of positive and negative entropies has been also discussed for gravastars\cite{Volovik2024} -- black holes with de Sitter interior.\cite{Dymnikova2020,MazurMottola2023,Mottola2025}  In this case one has the positive Bekenstein-Hawking entropy of the black hole horizon, and the negative entropy of the contracting de Sitter with negative Hubble parameter inside the black hole, $H<0$ in Eq.(\ref{G}). If the contracting de Sitter state occupies the entire interior volume inside the black hole horizon, $V=V_{\rm bh}=V_H$, the two horizons are annihilated and their entropies cancel each other. In the other presentation, the negative entropy of the volume cancels the positive surface entropy of the black hole horizon:
 \begin{equation}
S_{\rm black \, gravastar} = - V_H s_{\rm dS} + S_{\rm bh}= -\frac{A}{4G} +\frac{A}{4G}=0\,.
\label{gravastar}
\end{equation}
Zero entropy of this state correlates with the absence of Hawking radiation in this extremal limit.
A similar vanishing of the Hawking radiation temperature in the limit where the two horizons merge occurs in the holonomy corrected quantum gravity.\cite{Alonso-Bardaji2024,Alonso-Bardaji2025} 

For the white gravastar with the expanding de Sitter core the situation is opposite, but again with zero entropy in the extreme limit:
 \begin{equation}
S_{\rm white \,gravastar} = V_H s_{\rm dS} +S_{\rm wh}= \frac{A}{4G} -\frac{A}{4G}=0\,.
\label{gravastar}
\end{equation}

  \section{Discussion}
\label{DiscussionSec}

The de Sitter state has a special symmetry: it is homogeneous, and its curvature is constant in space. Since all the points in the de Sitter space are equivalent, this state is described by
local thermodynamics. This state has the local temperature, the local entropy density and the local energy density. The de Sitter thermodynamics also contains the gravitational degrees. These are described by a pair of the thermodynamically conjugate variables, the scalar curvature ${\cal R}$ and the effective gravitational coupling $K$. This pair of non-extensive variables is similar to a pair of non-extensive electrodynamic variables in the thermodynamics of dielectrics: the electric field ${\bf E}$ and the electric induction ${\bf D}$. Being non-extensive, variables $K$ and ${\cal R}$ naturally enter into the modified pressure, $P= p_{\rm vac}+ K{\cal R}$, thus providing the consistent first law of thermodynamics for the de Sitter state. 
 
 All these thermodynamic quantities are local and thus at first glance have no relation to the cosmological horizon. However, there are many connections with the physics of horizon. In particular, the local temperature is exactly twice the Gibbons-Hawking temperature, which enters the rate of the Hawking radiation from the horizon. Actually, there are two sides of the de Sitter thermodynamics, local and global. The latter refers to  the thermodynamics of the Hubble volume, and thus is related to the cosmological horizon. There is the holographic connection between these two sides: the entropy density integrated over the Hubble volume coincides with the entropy of the horizon, $S_{\rm Hubble}=S_{\rm horizon}=A/4G$. This also concerns the first law of thermodynamics: on one hand, due to the de Sitter symmetry the first law is valid for an arbitrary volume $V$ in Eq.(\ref{TdS}); on the other hand the first law is also applicable to the cosmological horizon in Eq.(\ref{FirstLaw1}). Note, that in both cases the temperature is the same, $T=H/\pi$. This consideration was also applied to the contracting de Sitter, for which one has $S_{\rm Hubble}=S_{\rm horizon}=-A/4G$. The entropy of contracting de Sitter is negative, since its horizon is similar to the horizon of white hole.

It would be interesting to extend the consideration to the apparent horizon.\cite{Odintsov2025b,Paul2025}

  {\bf Acknowledgements}.  I thank Dmitrii Diakonov and Sergei Odintsov for correspondence.


\begin{thebibliography}{999}

\bibitem{GibbonsHawking1977}
G. Gibbons and S. Hawking,
Cosmological event horizons, thermodynamics, and particle creation,
Phys. Rev. D {\bf 15}, 2738 (1977).

\bibitem{Paddy2002}
T. Padmanabhan, 
Classical and quantum thermodynamics of horizons in spherically symmetric spacetimes, 
Class. Quantum Grav. {\bf 19}, 5387--5408 (2002).

\bibitem{Odintsov2004}
 I.H. Brevik, S. Nojiri, S.D. Odintsov and L. Vanzo,
Entropy and universality of Cardy-Verlinde formula in dark energy universe,
Phys. Rev. D {\bf 70}, 043520 (2004),
arXiv:hep-th/0401073 [hep-th].

\bibitem{Markkanen2018}
Tommi Markkanen,
De Sitter Stability and Coarse Graining,
Eur. Phys. J. C {\bf 78}, 97 (2018),
arXiv:1703.06898 [gr-qc].

\bibitem{Akhmedov2020}
E.T. Akhmedov, K.V. Bazarov, D.V. Diakonov and U. Moschella,
Quantum fields in the static de Sitter universe,
Phys. Rev. D {\bf 102}, 085003 (2020).

 \bibitem{Galante2023} 
Damian A. Galante,
Modave Lecture Notes on de Sitter Space and Holography,
PoS(Modave2022)003,
arXiv:2306.10141.

\bibitem{Jacobson2023} 
Batoul Banihashemi, Ted Jacobson, Andrew Svesko, and Manus Visser,
The minus sign in the first law of de Sitter horizons,
JHEP 01:054, 2023, 
arXiv:2208.11706.

\bibitem{Diakonov2025}
D.V. Diakonov,
De Sitter entropy: on-shell versus off-shell,
arXiv:2504.01942 [hep-th].

\bibitem{Hooft2024} 
Gerard ’t Hooft, 
Alternative theory for the quantum black hole and the temperature of its quantum radiation,
chapter 3 in the book: The Black Hole Information Paradox: A Fifty-Year Journey, A. Akil and C. Bambi, editors,
arXiv:2410.16891.
 
 \bibitem{Painleve} 
P. Painlev\'e, 
La m\'ecanique classique et la th\'eorie de la relativit\'e, 
 C. R. Acad. Sci. (Paris) {\bf 173} , 677 (1921).

 \bibitem{Gullstrand} 
A. Gullstrand,
Allgemeine L\"osung des statischen Eink\"orper-problems in der Einsteinschen Gravitations-theorie,
Arkiv. Mat. Astron. Fys. {\bf 16}, 1-15 (1922).

\bibitem{Volovik2009} 
G.E. Volovik,  
Particle decay in de Sitter spacetime via quantum tunneling,
Pis'ma ZhETF {\bf 90}, 3--6 (2009); 
JETP Lett. {\bf 90}, 1--4 (2009);
arXiv:0905.4639 [gr-qc].

\bibitem{Volovik2025} 
G.E. Volovik, 
From Landau two-fluid model to de Sitter Universe,
to be published in Uspekhi,
https://ufn.ru/en/articles/accepted/39891/ ,
arXiv:2410.04392.

\bibitem{Maxfield2022} 
Henry Maxfield and Zahra Zahraee,
Holographic solar systems and hydrogen atoms: non-relativistic physics in AdS and its CFT dual,
JHEP 11 (2022) 093.

\bibitem{Bros2008} 
Jacques Bros, Henri Epstein and Ugo Moschella,
Lifetime of a massive particle in a de Sitter universe,
JCAP0802:003,2008,
arXiv:hep-th/0612184.

\bibitem{Polyakov2011} 
Dmitry Krotov, Alexander M. Polyakov,
Infrared Sensitivity of Unstable Vacua,
Nucl. Phys. B {\bf 849}, 410--432 (2011) ,
arXiv:1012.2107 [hep-th].


\bibitem{Volovik2024} 
G.E. Volovik, 
Thermodynamics and decay of de Sitter vacuum,
Symmetry {\bf 16}, 763 (2024).

\bibitem{Polyakov2010} 
A.M. Polyakov,
Decay of Vacuum Energy,
Nucl. Phys. B {\bf 834}, 316--329 (2010),
arXiv:0912.5503 [hep-th].

\bibitem{Volovik2024c} 
G.E. Volovik, 
de Sitter local thermodynamics in $f(R)$ gravity,
Pis’ma v ZhETF {\bf 119}, 560--561 (2024),
JETP Letters {\bf 119}, 564--571 (2024),
DOI: 10.1134/S0021364024600526,
arXiv:2312.02292v8.

\bibitem{Zeldovich1962} 
Ya.B. Zel'dovich,
The equation of state at ultrahigh densities and its relativistic limitations,
JETP {\bf 14}, 1143 (1962).

\bibitem{KlinkhamerVolovik2008} 
F.R. Klinkhamer and G.E. Volovik,  
$f(R)$ cosmology from $q$--theory, 
Pis'ma ZhETF {\bf 88}, 339--344 (2008);
JETP Lett. {\bf 88}, 289--294 (2008);
arXiv:0807.3896 [gr-qc].

\bibitem{Volovik2024d} 
G.E. Volovik, 
Thermodynamics of Einstein static Universe with boundary,
arXiv:2410.10549.

\bibitem{Odintsov2025} 
Shin'ichi Nojiri, Sergei D. Odintsov, Tanmoy Paul, Soumitra SenGupta,
Modified gravity as entropic cosmology,
arXiv:2503.19056.

\bibitem{Odintsov2025a} 
Sergei D. Odintsov, Simone D'Onofrio, Tanmoy Paul,
Generalized entropic dark energy with spatial curvature,
arXiv:2504.03470.

\bibitem{Odintsov2002}
M. Cvetic, S. Nojiri and S.D. Odintsov,
Black hole thermodynamics and negative entropy in de Sitter and anti-de Sitter Einstein-Gauss-Bonnet gravity,
Nucl. Phys. B {\bf 628}, 295--330 (2002),
arXiv:hep-th/0112045 [hep-th].

\bibitem{Volovik2020} 
G.E. Volovik,
Varying Newton constant and black hole to white hole quantum tunneling,
Universe {\bf 6}, 133 (2020),
arXiv:2003.10331.

\bibitem{Dymnikova2020} 
I. Dymnikova,
The fundamental roles of the de Sitter vacuum,
Universe {\bf 6}, 101  (2020).

\bibitem{MazurMottola2023} 
Pawel O. Mazur, Emil Mottola,
Gravitational Condensate Stars: An Alternative to Black Holes
Universe {\bf 9}, 88 (2023),
arXiv:gr-qc/0109035.


\bibitem{Mottola2025} 
Emil Mottola,
Gravitational Vacuum Condensate Stars in the Effective Theory of Gravity,
arXiv:2502.02519.

\bibitem{Alonso-Bardaji2024} 
 Asier Alonso-Bardaji,
 Formation of nonsingular spherical black holes with holonomy corrections
arXiv:2410.20529 [gr-qc].

\bibitem{Alonso-Bardaji2025} 
 Asier Alonso-Bardaji, David Brizuela, Marc Schneider,
Radiative properties of a nonsingular black hole: Hawking radiation and gray-body factor,
arXiv:2504.13050 [gr-qc].

\bibitem{Odintsov2025b} 
Sergei D. Odintsov, Tanmoy Paul, Soumitra SenGupta,
Natural validation of the second law of thermodynamics in cosmology,
Phys. Rev. D {\bf 111}, 043544 (2025), 
arXiv:2409.05009 [gr-qc].

\bibitem{Paul2025} 
Tanmoy Paul,
Origin of bulk viscosity in cosmology and its thermodynamic implications,
arXiv:2504.00422 [gr-qc]. 


\end{thebibliography}
\end{document}